\documentclass[12pt,preprint]{aastex}

\newcommand{\htco}{H$_2$CO}
\newcommand{\kms}{km\,s$^{-1}$}

\shorttitle{Formaldehyde as a Tracer of Extragalactic Molecular Gas I}
\shortauthors{M\"uhle et al.}

\begin{document}

\title{Formaldehyde as a Tracer of Extragalactic Molecular Gas \\ I. Para-\htco\ Emission From M\,82\footnote{Based on observations carried out with the IRAM 30-meter telescope. IRAM is supported by INSU/CNRS (France), MPG (Germany) and IGN (Spain).}}

\author{S. M\"uhle and E. R. Seaquist}
\affil{Department of Astronomy and Astrophysics, University of Toronto, 
       50 St. George Street, Toronto, ON M5S 3H4, Canada}
\email{muehle@astro.utoronto.ca, seaquist@astro.utoronto.ca}

\and

\author{C. Henkel}
\affil{Max-Planck-Institut f\"ur Radioastronomie, Auf dem H\"ugel 69, 
             D-53121 Bonn, Germany}
\email{p220hen@mpifr-bonn.mpg.de}

\begin{abstract}
Using the IRAM 30-m telescope and the 15-m JCMT, we explore the value of 
para-formaldehyde (p-H$_2$CO) as a tracer of density and 
temperature of the molecular gas in external galaxies. The target of
our observations are the lobes of the molecular ring around the center of the 
nearby prototypical starburst galaxy M\,82. 
It is shown that p-\htco\ provides one of the rare direct molecular 
thermometers. Reproducing the measured line intensities with a large 
velocity gradient (LVG) model, we find densities of 
$n_{\rm H2} \sim 7 \times 10^3\,{\rm cm^{-3}}$ and kinetic temperatures of 
$T_{\rm kin} \sim 200$\,K. The derived kinetic temperature is significantly 
higher than the dust temperature or the temperature deduced from ammonia 
(NH$_3$) lines, but our results agree well with the properties of the 
high-excitation component seen in CO.   
We also present the serendipitous discovery of the $4_2 \to 3_1$ line of 
methanol (CH$_3$OH) in the northeastern lobe, which shows --- unlike CO and 
\htco\ --- significantly different line intensities in the two lobes. 
\end{abstract}


\section{INTRODUCTION}

Molecular gas is regarded as the fuel for star formation. Increasing 
evidence for non-standard initial mass functions 
\citep[e.g.][]{paumard06,klessen07}, and warm molecular gas in starburst 
galaxies \citep[e.g.][]{rigopoulou02,mauers03} suggest that its 
physical properties may influence the star formation rate and the 
properties of the next generation of stars. 
Unfortunately, the physical properties of the molecular gas in external 
galaxies, in particular the kinetic temperature, are often not well 
constrained. The easily thermalized and optically thick CO $J = 1 \to 0$ and 
$2 \to 1$ transitions could constitute a good temperature tracer, 
but the filling factor of extragalactic clouds is poorly constrained.
Other commonly observed molecules like HCN and HCO$^+$ are good density 
tracers, but require an a priori knowledge of the kinetic temperature.
The inversion lines of the symmetric top molecule ammonia (NH$_3$) are 
frequently used as the galactic ``standard cloud 
thermometer''. However, in the disk of the Milky Way, the fractional abundance 
of NH$_3$ varies between $10^{-5}$ in hot cores \citep{mauers87} and $10^{-8}$ 
in dark clouds 
\citep{benson83}. Thus, ammonia may trace preferentially a specific component 
of the molecular gas and the assumption of an approximately constant 
fractional abundance on linear scales of a few 100\,pc is likely not valid. 
Other symmetric or slightly asymmetric top molecules may therefore be more 
favorable for extragalactic line studies.

\subsection{Formaldehyde as a Molecular Gas Tracer}
\label{h2co}

In this paper, we investigate the diagnostic properties of para-formaldehyde 
(p-\htco) lines in extragalactic sources. 
Formaldehyde is formed on the surface of dust grains by successive 
hydrogenation of CO \citep{watanabe02}, released into the gas phase by shocks 
or UV heating and subsequently destroyed by gas-phase processes. 
Variations in fractional abundance rarely exceed one order of magnitude. 
\citet{johnstone03} found only little variation in a variety of galactic 
environments ranging from cool protostellar candidates to hot photon-dominated 
regions and infrared sources with the latter showing the highest
discrepancy, by a factor of 5 to 10. These higher abundances are likely 
caused by the release of \htco\ molecules from grain surfaces in regions of 
massive star formation. 
As another example for the level of variability in the abundance of 
formaldehyde versus that of ammonia in active environments, the \htco\ 
abundance is the same in the hot core and the compact ridge of Orion A, whereas
the NH$_3$ abundance in the core is two orders of magnitude higher than that 
in the ridge \citep{charnley92}. 

Para-formaldehyde, which is a subspecies of the slightly asymmetric top 
molecule \htco, has a rich mm- and submm-spectrum of transitions.
Since the relative populations of the $K_a$ ladders are generally governed by 
collisions, line ratios involving different $K_a$ ladders 
(inter-ladder ratios)
are generally good tracers of the kinetic temperature. With the temperature 
known, line ratios within a single $K_a$ ladder, the so-called intra-ladder 
ratios,
whose populations are mainly determined by collisional excitation and radiative
 de-excitation, sensitively probe the gas density \citep{mangum93}. 
The $K$-doublet transitions of ortho-\htco\ in the cm-range, 
$1_{10} \to 1_{11}$ at $\lambda = 6$\,cm and $2_{11} \to 2_{12}$ at 
$\lambda = 2$\,cm, have been 
observed in a number of external galaxies since the 1970s 
\citep[e.g.][]{gardner74,seaquist90}. \htco\ lines in the mm-range were 
detected in several galaxies including the LMC, M\,82 and NGC\,253 since the 
1990s \citep[e.g.][]{baan90,johansson94}, but to date there are only three 
published studies of the excitation conditions of extragalactic molecular gas 
that utilize multi-line \htco\ observations in combination with a non-LTE 
treatment of the radiative transfer
\citep{huette97,heikkilae99,wang04}.

\subsection{Target Selection: M\,82}

\begin{figure}
\plotone{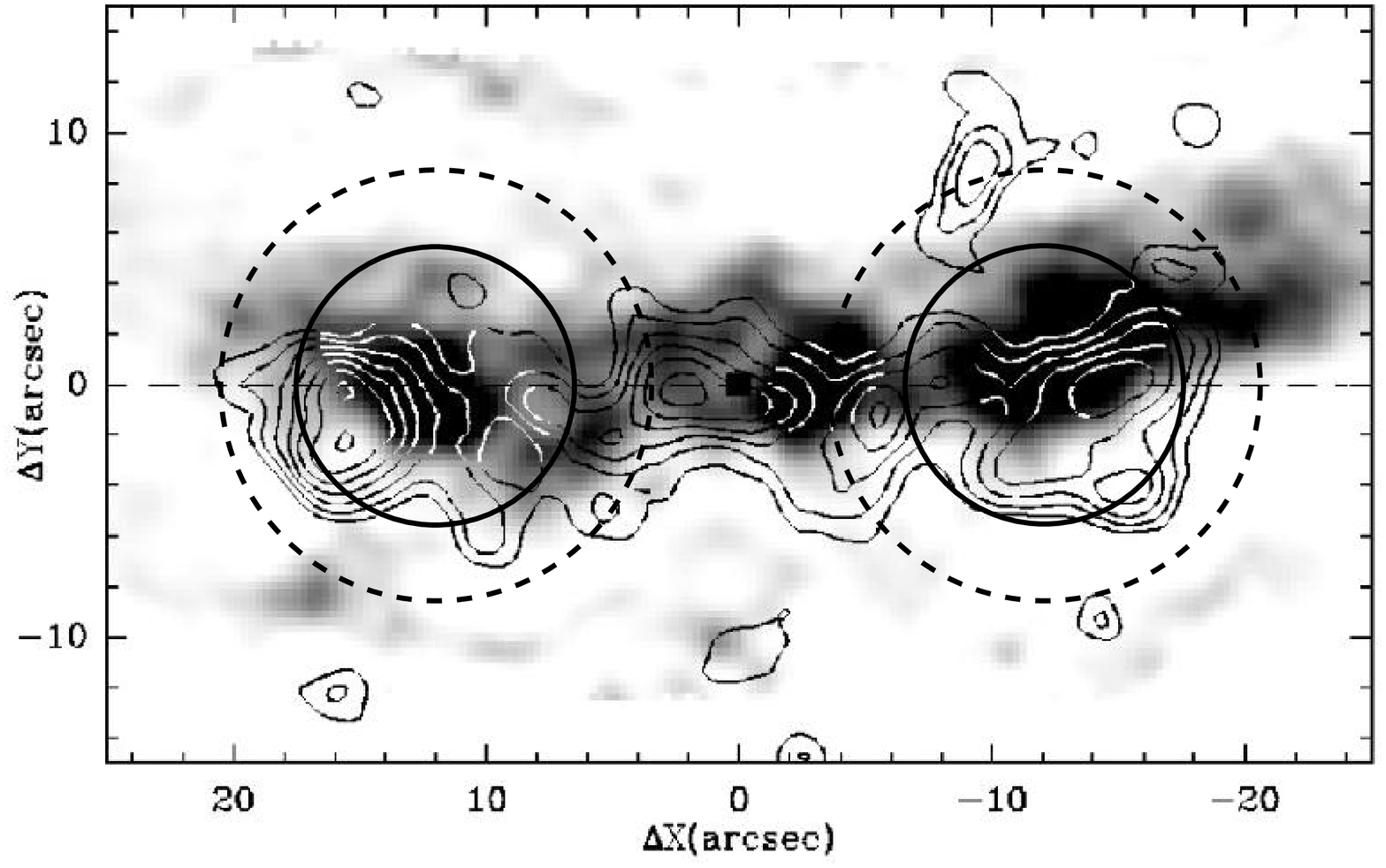}  
\caption{Contours of the HCO($F=2\to 1$) emission superposed on a 
high-resolution CO($2 \to 1$) map of the molecular ring in M\,82 adapted from 
\citet{garcia02}. Map offsets are relative to the dynamical center of the 
galaxy (${\rm \alpha_{2000}}=09^{\rm h}55^{\rm m}51\fs9$, 
${\rm \delta_{2000}}=69\degr40\arcmin47\farcs1$) with the x-axis along the 
plane of the molecular ring (P.A.$=70\degr$ east of north). 
The observed pointing positions and the corresponding beam widths are marked 
by the solid (11\arcsec\ at 218\,GHz) and the dashed (17\arcsec\ at 146\,GHz) 
circles.
\label{map}
}
\end{figure}

The target for our exploratory investigation is the nearby prototypical 
starburst galaxy M\,82. As the most prominent galaxy in the northern hemisphere
with a nuclear starburst, 
M\,82 has been the target of numerous molecular line studies 
\citep[e.g.][]{fuente06,seaquist06}.
Most of the excitation studies focus on the central 1\,kpc disk, where 
the molecular gas is concentrated in a circumnuclear ring around the center of 
the starburst  \citep[e.g.][]{garcia02,seaquist06} with the more highly 
excited lines found at smaller radii (Fig.~\ref{map}, see also \citet{mao00} 
for an overview of ring diameters). 
Note that the two lobes of the molecular ring in this highly inclined galaxy 
can be easily separated with single-dish telescopes and the observed molecular 
lines in the lobes have typical widths of only $100$ to $ 150$\,\kms\ 
\citep[e.g.][]{seaquist98,garcia02}, thus greatly reducing the potential for
line blending.
The most comprehensive CO line studies to date \citep{mao00,ward03} suggest 
that the 
broad data base of CO and $^{13}$CO lines observed towards the molecular lobes 
of M\,82 can not be fitted by a single-component large velocity gradient (LVG)
model and that the majority of the molecular gas may actually be in a 
high-excitation state. 
The \htco($1_{10} \to 1_{11}$) transition at $\lambda = 6$\,cm was detected 
towards the nucleus of M\,82 {\em in absorption} by \citet{graham78}, whereas 
the \htco($2_{11} \to 2_{12}$) line at $\lambda = 2$\,cm was observed 
{\em in emission}, which was 
interpreted as evidence for very dense ($10^6$\,cm$^{-3}$) molecular gas in 
the molecular ring \citep{baan90}. 
The latter study also reported the first detection of an \htco\ emission line 
in the mm-range in M\,82. A first multi-line \htco\ excitation study 
towards M\,82 
was performed by \citet[][see also Sect.~\ref{h2co}]{huette97} using an LVG 
model that included 16 ortho-\htco\ levels of the $K_a=1$ ladder. 

\subsection{Line Selection}
\label{select}

Having developed a comprehensive LVG code for ortho-\htco\ 
(40 levels) and para-\htco\  (41 levels, see Sect.~\ref{lvg}), we searched for
the set of transition lines best suited for extragalactic observations while
trying to minimize observational uncertainties. The usually small filling 
factor of extragalactic observations and the decreased sensitivity of 
ground-based receivers at high frequencies limits the choice to the strongest 
lines, usually corresponding to transitions between low excitation levels. 
Among the \htco\ lines
that can be observed with present-day receivers, the para-\htco\ transitions
$3_{03} \to 2_{02}$, $3_{22} \to 2_{21}$ and $3_{21} \to 2_{20}$ stand out by 
being close in frequency and at the same time strong enough for extragalactic 
observations. With rest frequencies at 218.22\,GHz, 218.48\,GHz and 
218.76\,GHz, all three lines can be observed simultaneously, if the receiver 
and spectrometer both offer a bandwidth of 1\,GHz. That way, the observed 
inter-ladder 
line ratio \htco($3_{03} \to 2_{02}$)/\htco($3_{21} \to 2_{20}$), which is a 
good temperature tracer, will be free from uncertainties related to 
pointing accuracy, calibration issues or different beam widths. Due to the 
large line widths of extragalactic observations ($\geq 100$\,\kms), the 
\htco($3_{22} \to 2_{21}$) line may be blended with methanol emission (see 
Sect.~\ref{methanol}) and thus be of limited use for the LVG analysis. At the 
very high densities \citep[$10^6$\,cm$^{-3}$][]{baan90} and kinetic 
temperatures of 50\,K to 100\,K inferred for \htco\ in M\,82, our LVG code 
suggests a line ratio 
\htco($3_{21} \to 2_{20}$)/\htco($5_{24} \to 4_{23}$) of the order of unity, 
which sensitively traces the gas density. In addition, observing the 
\htco($3_{21} \to 2_{20}$) line with the IRAM 30-m telescope and the 
para-\htco($5_{24} \to 4_{23}$) line with the 15-m dish of the JCMT results in 
data with nearly the same beam width (11\arcsec\ vs.~13\arcsec) and thus 
minimizes the uncertainty inherent in line ratios derived from observations 
consisting of single pointings. Using a spectrometer with a bandwidth
of at least 900\,MHz, the \htco($5_{24} \to 4_{23}$) transition at 363.95\,GHz 
can be observed simultaneously with the weaker and heavily blended para-\htco\ 
transitions $5_{42} \to 4_{41}$ and $5_{41} \to 4_{40}$ at 364.10\,GHz as 
well as with the blended ortho-\htco\ transitions $5_{33} \to 4_{32}$ and
$5_{32} \to 4_{31}$ at 364.28\,GHz and 364.29\,GHz, respectively.
The non-detection of the \htco($5_{24} \to 4_{23}$) line 
(see Section~\ref{results}) hinted at a much lower
gas density regime than previously assumed and necessitated the addition of 
the relatively strong para-\htco($2_{02} \to 1_{01}$) line at 145.60\,GHz to 
the 
selected transitions in order to be able to derive the density-sensitive 
intra-ladder line ratio \htco($2_{02} \to 1_{01}$)/\htco($3_{03} \to 2_{02}$). 
\\

Here we present the results of our para-\htco\ observations. To our knowledge, 
this is the first dedicated search for para-\htco\ transition lines of the 
$K_a = 2$ ladder outside of the Galactic neighborhood (Milky Way and Magellanic
Clouds). The distribution
of the \htco\ emission and its correlation to other tracers will be discussed 
in a forthcoming paper based on high-resolution ($\sim 4$\arcsec) ortho-\htco\
data recently obtained with the VLA. 
After a description of the observations and the data reduction strategy in 
Sect.~2, we present the observational results in Sect.~3. In Sect.~4, we 
describe our LVG code and the parameter space covered,
before we discuss our results and compare them to other molecular excitation 
studies. Finally, we report the serendipitous detection of methanol emission 
in one of the lobes
(Sect.~5). Our conclusions are summarized in Sect.~6.

\section{OBSERVATIONS AND DATA REDUCTION}

For this study, we selected the para-\htco\ transitions described in 
Sect.~\ref{select} to be observed at 
nearly the same spatial resolution with the IRAM 30-m telescope and the 15-m 
JCMT. The parameters of the observations are summarized in Table~\ref{obs}.

\subsection{Observations with the IRAM 30-m Telescope}
\label{iram}

\begin{figure}
\plotone{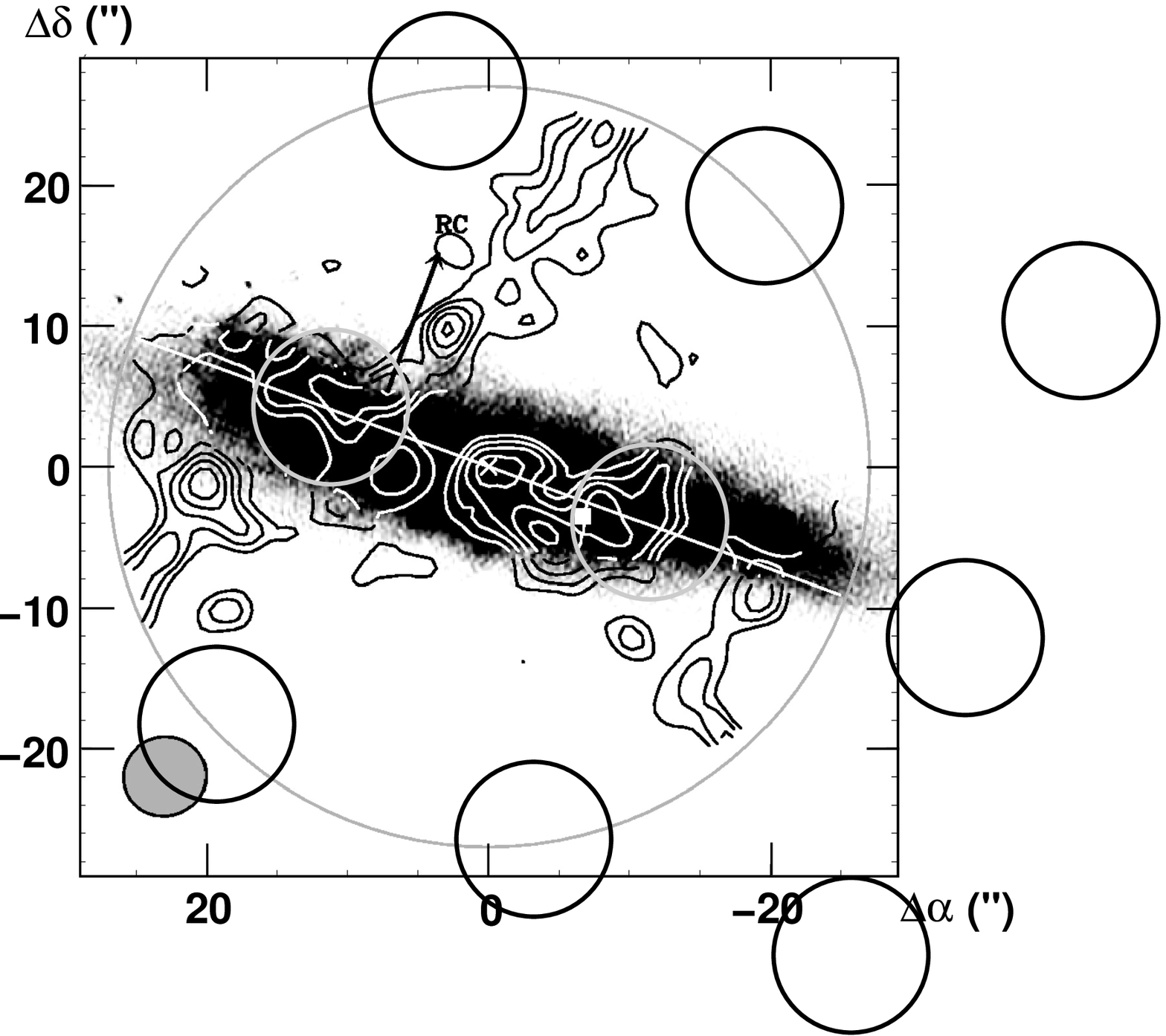}  
\caption{Contours of the velocity-integrated SiO($v=0$, $J=2-1$) emission 
superposed on a radio continuum emission image at 4.8\,GHz, adapted from 
\citet{garcia01}. Map offsets are relative to the dynamical center of the 
galaxy (${\rm \alpha_{2000}}=09^{\rm h}55^{\rm m}51\fs9$, 
${\rm \delta_{2000}}=69\degr40\arcmin47\farcs1$). The large circle delimits the
primary beam field of the SiO interferometric observations at 87\,GHz 
(55\arcsec), while the synthesized beam is shown in the bottom left corner. The
white square marks the position of SNR 41.95+57.5. The radio continuum filament
is indicated by an arrow. The pointing positions observed with the HERA array 
and the beam width of 11\arcsec\ are marked by the gray and black solid
circles, where the gray circles cover the molecular lobes in M\,82.
\label{siomap}
}
\end{figure}

For the observations at 218\,GHz, the IRAM 30-m telescope with the heterodyne 
receiver array HERA, consisting of nine dual-polarization receivers arranged 
in form of a center-filled square with a pixel separation of 24\arcsec, was 
pointed towards the southwestern lobe (SW lobe) of M\,82 at 
${\rm \alpha_{2000}}=09^{\rm h}55^{\rm m}49\fs4$, 
${\rm \delta_{2000}}=69\degr40\arcmin43\farcs1$ during four nights in 
March/April 2005.  The derotator optical assembly was set to keep the pixel 
pattern stationary in equatorial coordinates at an angle of 20\degr\ clockwise
relative to the axis of right ascension, so that the central row of pixels was 
aligned with the major axis of the molecular ring with the two molecular lobes 
being covered by two adjacent pixels (Fig.~\ref{map}) and the third pixel 
pointing at a position 24\arcsec\ southwest of the SW lobe. 
In this arrangement, the remaining pixels of the array pointed towards six 
positions 24\arcsec\ northwest and southeast of the major axis 
(Fig.~\ref{siomap}). The spectra were obtained in a wobbler switching mode 
with a beam throw of 240\arcsec\ in azimuth.
As backend, we used the WILMA autocorrelator, which provides a bandwidth of 
1024\,MHz (512 channels) in each of its 18 units. Having tuned the 
receivers to a rest frequency of 218.48\,GHz at a velocity of 200\,\kms, 
each bandpass thus covered simultaneously the para-\htco($3_{03} \to 2_{02}$), 
($3_{22} \to 2_{21}$) and ($3_{21} \to 2_{20}$) transitions
(see Table~\ref{obs}). Good winter weather conditions 
resulted in system temperatures of 200 to 400\,K ($T_A^*$) and a pointing 
accuracy of 3\arcsec\ or better. The chopper wheel calibration was checked by 
observing the C$^{18}$O($2 \to 1$) line at 219.56\,GHz toward the reference 
source CLR\,2688 and the relative stability of each tuning was determined by 
observing the \htco\ emission of 
NGC\,2264\,IR and DR\,21\,OH. Both checks suggest a calibration uncertainty of 
less than 15\%. The temperature scale was converted to main-beam 
brightness temperature ($T_{\rm mb}$) adopting a main beam efficiency of 
$B_{\rm eff}=0.55$ and a forward efficiency of $F_{\rm eff}= 0.90$. 

The para-\htco($2_{02} \to 1_{01}$) spectra at 145.60\,GHz were obtained
at the IRAM 30-m telescope during two days in August 2006. Using the wobbler 
switching mode (200\arcsec\ throw) of the new control system 
(NCS), we observed each molecular lobe in M\,82 with receivers C150 and D150 
tuned to the same frequency and each connected to two parts of the  1\,MHz 
filterbank, resulting in a bandwidth of 512\,MHz with a channel spacing of 
1\,MHz for each receiver. 
During the observations, the system temperature was 
200 to 350\,K and the pointing was accurate within 3\arcsec\ to 
4\arcsec. Daily observations of the galactic sources W51E1 and/or W3OH suggest 
a calibration 
uncertainty of 10\% or less. For the conversion of the antenna temperature to 
main-beam brightness temperature ($T_{\rm mb}$) a main beam efficiency of 
$B_{\rm eff}=0.69$ and a forward efficiency of $F_{\rm eff}= 0.93$ were 
adopted.

\subsection{Observations with the JCMT}
\label{jcmt}

During four shifts in December 2004 and January 2005, the SW lobe of M\,82 was 
observed with the dual-mixer receiver B3 at the JCMT\footnote{Program ID M04BC10; the James Clerk Maxwell Telescope is operated by The Joint Astronomy Centre on behalf of the Science and Technology Facilities Council of the United Kingdom, the Netherlands Organisation for Scientific Research, and the National Research Council of Canada.} tuned to a rest frequency of 364.1\,GHz and a velocity of 180\,\kms. The observations were 
carried out in beam switching mode with a throw of 210\arcsec. The DAS 
autocorrelator provided a bandwidth of 920\,MHz with a spectral channel 
separation of 1.25\,MHz in each of its two receiver channels, sufficient to 
cover simultaneously 
the para-\htco\ transitions ($5_{24} \to 4_{23}$),  ($5_{41} \to 4_{40}$) and  
($5_{42} \to 4_{41}$) as well as the ortho-\htco\ transitions 
($5_{33} \to 4_{32}$) and ($5_{32} \to 4_{31}$) (see Table~\ref{obs}). The 
system temperatures during the four shifts varied from $\sim 750$\,K to 
$\sim$1650\,K and the uncertainty of the chopper wheel calibration was about 
15\% as indicated by observations of the standard source IRC+10216. 
The temperature scale was converted to main-beam 
brightness temperature using a beam efficiency of $B_{\rm eff}=0.63$.

\subsection{Data Reduction}
\label{reduction}

All data were reduced, converted to the main-beam temperature scale and 
analyzed using the GILDAS software package CLASS. 
Each individual spectrum was inspected 
for bad channels, standing waves or other baseline problems before a linear 
baseline was subtracted. The spectra at each position were then weighted 
according to their noise level, combined and smoothed to a common 
velocity resolution of 10\,\kms. The data obtained at 218\,GHz with the newly 
commissioned WILMA 
backend required special attention, because a large fraction of the individual
spectra showed baseline problems ranging from prominent DC offsets to subtle 
ripples in certain sections of the bandpass. 
As a first step, the spectra were 
sorted into different groups according to their baseline quality.
In order to guard against more subtle baseline problems compromising the 
determination of line strengths, we then fitted the lines in each individual 
spectrum using the multi-line fit of the corresponding combined spectrum 
(Section~\ref{results}) as a template. When we selected only the spectra with 
the best baseline quality, 
representing $\sim 40$\% of the data, the $\chi^2$-tests yielded an 
agreement 
between the distribution of the fitted intensities and the expected Gaussian 
distribution at a 95\% confidence level for each line in the spectra. The 
uncertainty of each fitted line intensity shown in Table~\ref{lines} is the 
uncertainty in the mean
of the Gaussian distribution $\sigma_{<I>} = \sigma_I / \sqrt{N}$ with  
$\sigma_I$ the standard deviation of the Gaussian distribution and $N$ the 
number of selected spectra.

\section{RESULTS}
\label{results}

\begin{figure}
\plotone{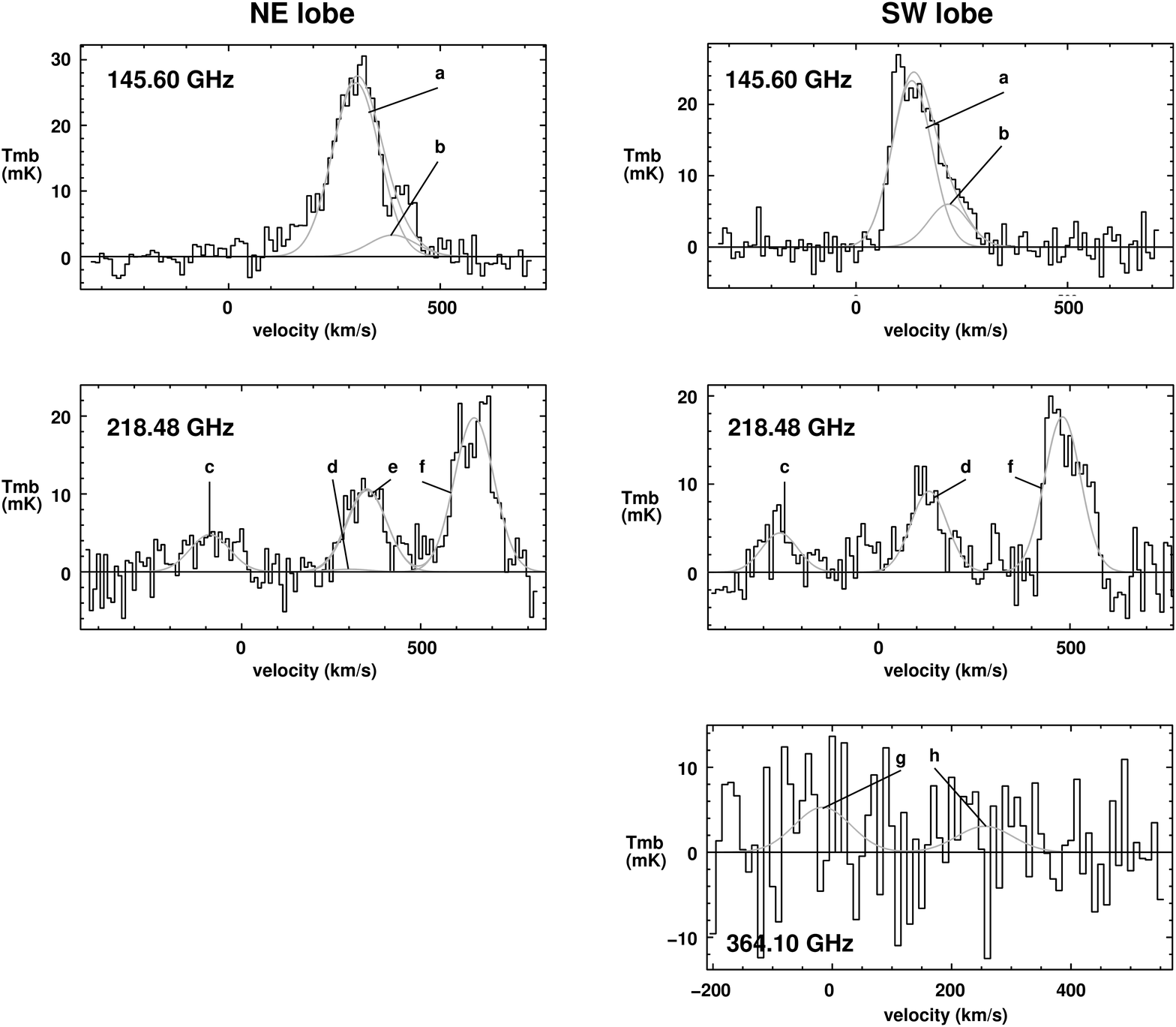} 
\caption{Observed spectra of the NE (left) and the SW lobe (right) of 
M\,82. Each spectrum is labeled with the frequency the receiver was tuned to.
Thus, the velocity scale of the 218.48\,GHz spectra refers to the 
\htco($3_{22} \to 2_{21}$) line. The \htco($3_{03} \to 2_{02}$) and the 
\htco($3_{21} \to 2_{20}$) transitions are offset by 348.5\,\kms\ and 
$-389.7$\,\kms, respectively, while the CH$_3$OH($4_2 \to 3_1\, E$) line is 
offset by 49.4\,\kms.  In the 146\,GHz spectra, the HC$_3$N($16 \to 15$) line 
is offset by  86.46\,\kms\ from the \htco($2_{02} \to 1_{01}$) emission.
Gaussian line profiles were fitted to every individual line assuming 
the same line width and velocity for all lines. All identified lines of a 
spectrum were fitted simultaneously with these constraints. The 
curves show the Gaussian fit to each individual line as well as the spectrum 
resulting from a superposition of all identified lines: a = \htco($2_{02} \to 1_{01}$), b = HC$_3$N($16 \to 15$), c = \htco($3_{21} \to 2_{20}$), d = \htco($3_{22} \to 2_{21}$), e = CH$_3$OH($4_2 \to 3_1\, E$), f = \htco($3_{03} \to 2_{02}$), g = \htco($5_{33} \to 4_{32}$) + \htco($5_{32} \to 4_{31}$), h = \htco($5_{24} \to 4_{23}$).  
\label{spectra}
}
\end{figure}

We have detected the \htco($2_{02} \to 1_{01}$) line at 146\,GHz as well as the
transitions \htco($3_{03} \to 2_{02}$), \htco($3_{22} \to 2_{21}$) and 
\htco($3_{21} \to 2_{20}$) at 218\,GHz in both molecular lobes of M\,82. This 
constitutes the 
first detection of \htco($3_{22} \to 2_{21}$) emission from an external galaxy.
We did not detect any \htco\ lines at 364\,GHz, probably because of the limited
sensitivity of our observations. The spectra of the molecular lobes and the 
fitted Gaussian lines are displayed in Fig.~\ref{spectra}. 
The basic assumption that the observed lines are all emitted from the same 
volume of molecular gas with the same average physical properties implies that 
all lines should have the same velocity profile, parametrized in a Gaussian fit
by the central velocity and the line width. Therefore we fixed the 
velocity and line width of all spectral features to the values derived 
from simultaneously fitting all the lines in the 218\,GHz spectra. The derived 
velocities and widths in the two lobes are in very good agreement 
with the results of a fit to the strong \htco($3_{03} \to 2_{02}$) line alone. 
The parameters of the Gaussian fits and the resulting integrated intensities 
are summarized in Table~\ref{lines}.

The 218\,GHz spectrum of the SW lobe is reproduced well by the three 
\htco\ lines near this frequency. The central velocity is 132\,\kms\ (local 
standard of rest, hereafter LSR); the line width is 111\,\kms. In the spectrum 
of the NE lobe, the central velocity is 301\,\kms\ (LSR) and the line width is 
131\,\kms. The apparent shift of the \htco($3_{22} \to 2_{21}$) line relative 
to the central velocity indicates the presence of at least one additional 
emission line, which we identify as the methanol CH$_3$OH($4_2 \to 3_1\, E$) 
transition 
at 218.44\,GHz. The ratio of the fitted intensities of the highly 
blended lines \htco($3_{22} \to 2_{21}$) and CH$_3$OH($4_2 \to 3_1\, E$) can
vary strongly with slight changes in the central velocity or the line width, 
which suggests that the fitting procedure is strongly influenced by the noise 
in the spectra. In addition, there might be faint HC$_3$N($24 \to 23$) 
emission at 218.33\,GHz in the spectrum of the NE lobe. However, the effect of 
such an additional component 
on the other fitted Gaussians is well inside the observational errors and
can be neglected. The integrated line
intensities resulting from a simultaneous fit of all five emission lines are 
given in parentheses in Table~\ref{lines}. A simultaneous fit of all five 
lines to the spectrum of the SW lobe gives a stringent upper limit to 
the HC$_3$N($24 \to 23$) emission at this position. 

\begin{figure}
\plottwo{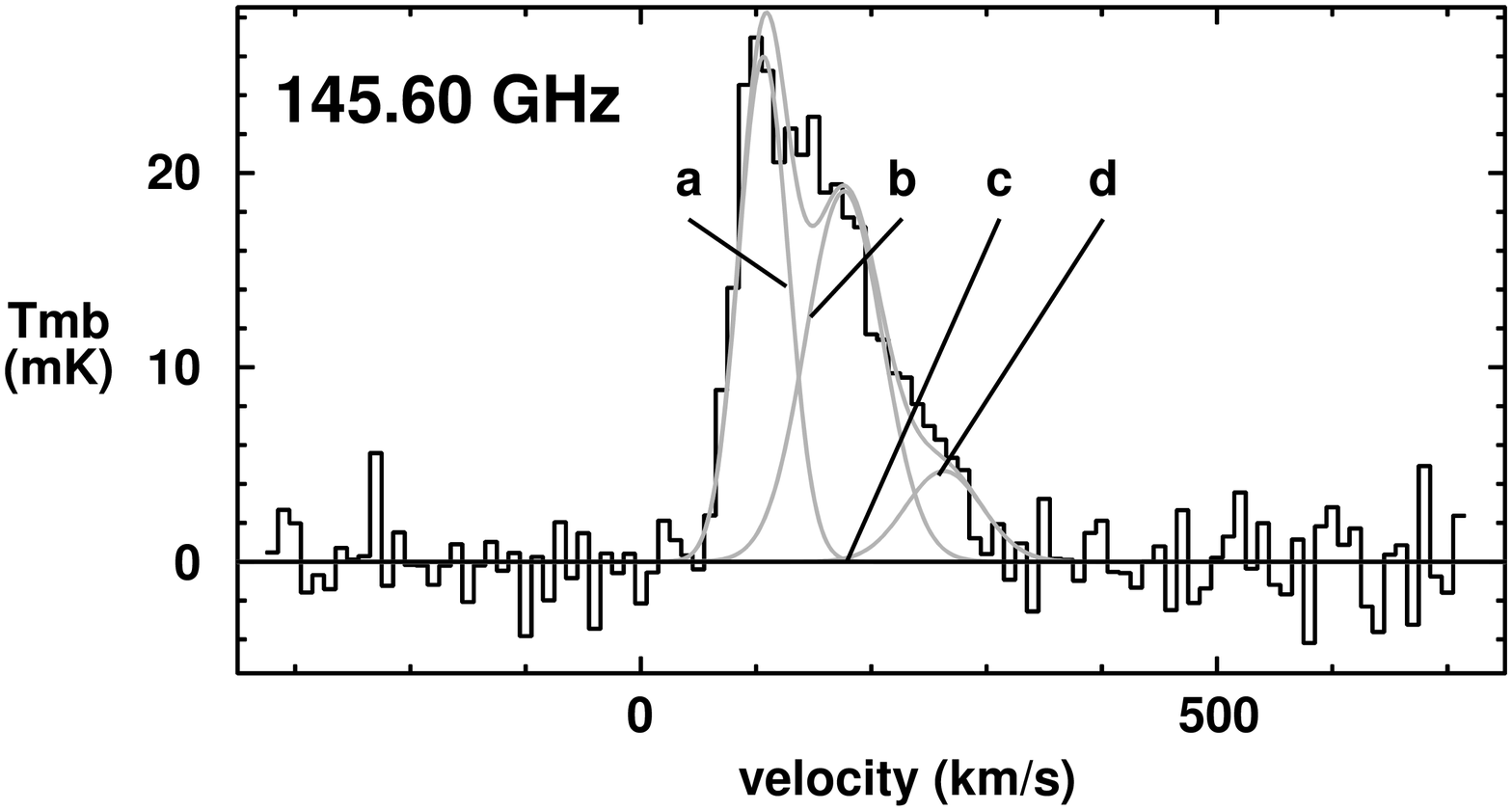}{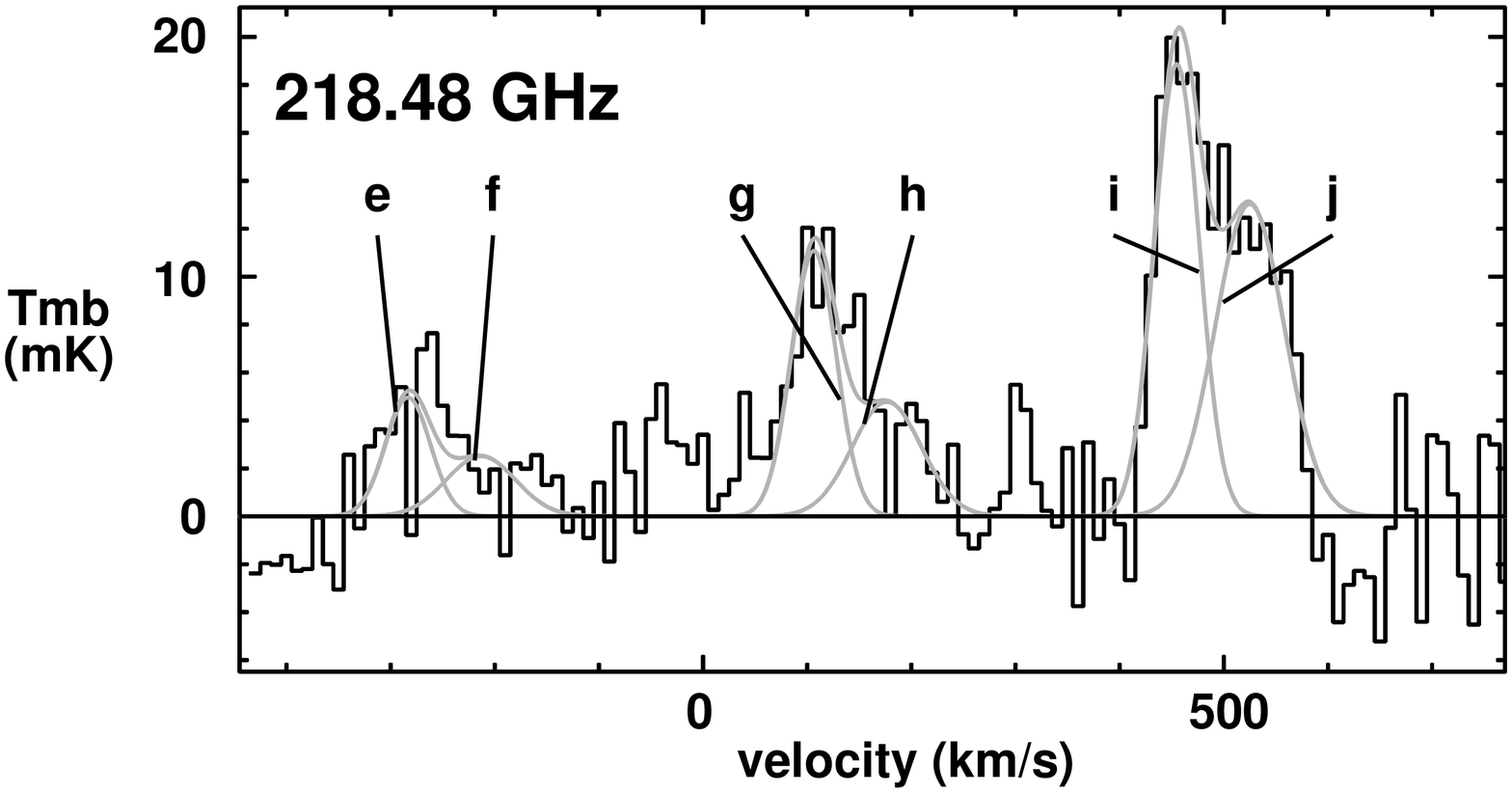} 
\caption{Spectra of the SW lobe observed at 146\,GHz (left) and 218\,GHz 
(right). To better match the observed line profiles, two velocity components 
($v_1$ and $v_2$) with a mutual separation of 68\,\kms\ were fitted to 
each previously 
identified line. The curves show the Gaussian fit to each individual line as 
well as the spectrum resulting from a superposition of all identified lines: 
a = \htco($2_{02} \to 1_{01})_{v1}$, 
b = \htco($2_{02} \to 1_{01})_{v2}$, 
c = HC$_3$N($16 \to 15)_{v1}$, 
d = HC$_3$N($16 \to 15)_{v2}$,
e = \htco($3_{21} \to 2_{20})_{v1}$,
f = \htco($3_{21} \to 2_{20})_{v2}$, 
g = \htco($3_{22} \to 2_{21})_{v1}$, 
h = \htco($3_{22} \to 2_{21})_{v1}$,
i = \htco($3_{03} \to 2_{02})_{v1}$,
j = \htco($3_{03} \to 2_{02})_{v2}$. Note that the \htco($2_{02} \to 1_{01})_{v2}$ and the HC$_3$N($16 \to 15)_{v1}$ lines are strongly blended 
with a separation in velocity of only 16.5\,\kms.   
\label{twocomp}
}
\end{figure}

In the SW lobe, the two \htco\ lines ($3_{03} \to 2_{02}$) and 
($2_{02} \to 1_{01}$), detected with high signal-to-noise ratios, show a 
similar deviation from a Gaussian profile. They are skewed, with a 
blue-shifted peak and a weaker red line wing. The peak at 
$v \sim 100$\,\kms\ may originate from a gas component that has been 
identified previously in interferometric maps and may constitute the eastern 
part of the molecular super-shell \citep{weiss99}. To better account for these
line profiles, we  fitted two velocity components,  a ``narrow'' one at 
$\sim 100$\,\kms\ and a ``wide'' one, to each identified spectral line using 
the velocity and line width derived from the \htco($3_{03} \to 2_{02}$) line 
(Table~\ref{lines}, Fig.~\ref{twocomp}). While the line profiles of the two 
strongest lines are fitted reliably by those two components, the result for 
the much weaker \htco($3_{21} \to 2_{20}$) line may be dominated by noise. 

We also inspected the spectra obtained with the other pixels of the HERA array
in order to search for possible outflows. Filaments of SiO emission, another 
molecule related to dust-chemistry and shocks, have been found at a distance 
as large as 24\arcsec\ from the central plane (Fig.~\ref{siomap}). 
No significant \htco\ emission was found in any of these spectra above a noise 
level of $\sim 2$\,mK (on the T$_{\rm mb}$ scale). Note, however, that \htco\ 
filaments may be present in the halo at locations not covered by our 
observations. 

In both 146\,GHz spectra, the \htco($2_{02} \to 1_{01}$) line at 145.60\,GHz 
is partially 
blended with the HC$_3$N($16 \to 15$) line at 145.56\,GHz, but the two lines 
can easily be separated given the constraints derived from the 218\,GHz 
spectra (Table~\ref{lines}). We did not detect any \htco\ emission at 364\,GHz 
from 
the SW lobe. The values given in Table~\ref{lines} are 3$\sigma$ upper limits, 
derived from the rms noise with the assumption that these lines 
have the same width and central velocity as those at 218 and 146\,GHz.

\section{DISCUSSION}

\subsection{The Physical Conditions of the Dense Molecular Gas}
\label{lvg}

For the analysis of the derived integrated line intensities, we have developed 
a non-LTE code for para-\htco\ adopting the Large Velocity Gradient (LVG) 
approximation and choosing a spherically symmetric cloud geometry.
The choice of a particular cloud geometry can affect the resulting gas 
densities. In the LVG approximation, the escape probability $\beta_{ij}$ for a 
photon of a transition $ij$ is 
$\beta_{ij} = (1-\exp(-\tau_{ij}))\,\tau_{ij}^{-1}$ for a model with 
spherically symmetric geometry compared to 
$\beta_{ij} = (1-\exp(-3\tau_{ij}))\,(3\tau_{ij})^{-1}$ for a plane-parallel 
geometry. Thus, applying a plane-parallel 
instead of a spherical cloud geometry, photon trapping sets in at 
lower optical depths. In the optically thick case, the particle densities 
derived from the observed line ratios can be lower by up to half an order of 
magnitude \citep[e.g.][]{ward03}. Our code is based on the collision rates with
He by 
\citet{green91}, the collision rates with H$_2$ being approximated by a scaling
factor of 1.37 \citep{schoeier05}, and includes 41 para-\htco\ levels, up to 
210\,cm$^{-1}$ (300\,K) above the ground state. We checked our code 
successfully by 
using the physical conditions derived in this study as input to the on-line 
version of RADEX\footnote{http://www.strw.leidenuniv.nl/$\sim$moldata/}, a 
program using a similar non-LTE code, and by comparing its output with our 
observed and derived line ratios and optical depths. 

The velocity profiles of the two strong lines \htco($2_{02} \to 1_{01}$) and 
\htco($3_{03} \to 2_{02}$) are very similar in each lobe, suggesting that 
the \htco\ emission is likely confined to a region within the smallest beam of 
the observations (11\arcsec).  
Interferometric maps of the integrated intensities of molecular transitions 
like CO($2\to1$), HCO$^+(1\to0)$ and HCO($F=2\to1$) 
\citep[e.g.][]{seaquist98,weiss01b,garcia02} 
indicate that the NE and the SW lobes can be approximated by circular 
Gaussian distributions of 7\arcsec\ to 8\arcsec\ in diameter (FWHM). Here we 
adopt a  Gaussian source distribution with a size of $\theta_s=7\farcs5$, 
unless noted otherwise, and convolve all intensities to a common resolution of
17\arcsec.
We calculated the expected line ratios for a large number of physical 
conditions, covering the parameter space for a kinetic temperature of 
$T_{\rm kin}=5$ to 300\,K in steps of 5\,K, a molecular gas density of 
$\log{n_{\rm H2}}=3.0$ to 6.0 (in cm$^{-3}$) in steps of 0.1 and a para-\htco\ 
column density per velocity interval of $\log{N_{\rm pH2CO}/\Delta v}=10.5$ to
14.5 (in cm$^{-2}$\,km$^{-1}$\,s) in steps of 0.1. The assumed ambient 
radiation field is the cosmic background radiation field 
($T_{\rm bg}=2.73$\,K). Since the 
\htco($3_{22} \to 2_{21}$) line in the NE lobe is heavily blended with a 
nearby methanol line (see Section \ref{methanol}) and does not add significant 
constraints to the physical conditions, we do not include the 
line in our LVG analysis for either lobe. 

Comparing the computed line ratios with our observational result 
(Table~\ref{ratios}), we can 
significantly constrain the range of possible physical conditions in the 
molecular lobes of M\,82\footnote{Splitting the observed \htco\ line profiles 
in the SW lobe into two velocity components (Table~\ref{lines}),  
the \htco($2_{02} \to 1_{01}$)/\htco($3_{03} \to 2_{02}$) line ratio 
suggests that the narrow component has a higher density than the wide 
component 
at a given abundance per velocity gradient and kinetic temperature. However, 
since the other line ratios are highly uncertain because of the low 
signal-to-noise ratio in the \htco($3_{21} \to 2_{20}$) line, we do not treat 
the two velocity component separately in the following analysis.}. 
Note that in reality, the molecular gas is likely to be inhomogeneous in its 
properties such as temperature and density. 
However, given the limited number of measurements, we are constrained to 
consider only a one-component model, yielding the average properties of the 
molecular gas phase traced by \htco. 

The observed integrated line intensity of extragalactic sources is usually 
only a small fraction of the intensity predicted by the LVG calculations. This 
fraction $F = T_{\rm obs}/T_{\rm lvg}$, the so-called filling factor or 
dilution factor, is the product of several components $F = f_b\,f_a\,f_v$
and provides insights into the structure of the observed source. 
The beam filling factor $f_b = \theta_s^2\,(\theta_s^2+\theta_b^2)^{-1}$ 
accounts for the fact that the source does not extend over the whole beam area.
Assuming a source size of $\theta_s=7\farcs5$, $f_b = 0.163$ for a beam width 
of $\theta_b=17$\arcsec. At the distance of M\,82 of $D = 3.9$\,Mpc 
\citep{sakai99}, a 17\arcsec\ beam covers an area with a diameter of 320\,pc. 
Thus, it is reasonable
to assume that the observed emission does not originate from a single molecular
cloud, but rather represents an ensemble of giant molecular 
clouds, a giant molecular association. The individual clouds are expected to 
have (much) smaller radii than the overall source size and (much) smaller line 
widths than the observed beam-averaged width. Measures of this small-scale 
structure are the area filling factor $f_a = C\,r^2\,R_s^{-2}$ and the 
velocity filling factor $f_v = \Delta v_{\rm cloud}\,\Delta v_{\rm line}^{-1}$,
where $r$ and $\Delta v_{\rm cloud}$ are the radius and velocity width of an 
individual cloud, $R_s$ is the radius of the source, $\Delta v_{\rm line}$ is 
the observed line width, and $C$ is the total number of clouds in the beam. 
Since the latter two filling factors can not be separated, we combine them into
the small-scale dilution factor $F_{sc} = f_a\,f_v$. Note that $F_{sc}$ 
represents the volume filling factor of the velocity cube and thus can not be 
larger than unity.

\begin{figure}
\plotone{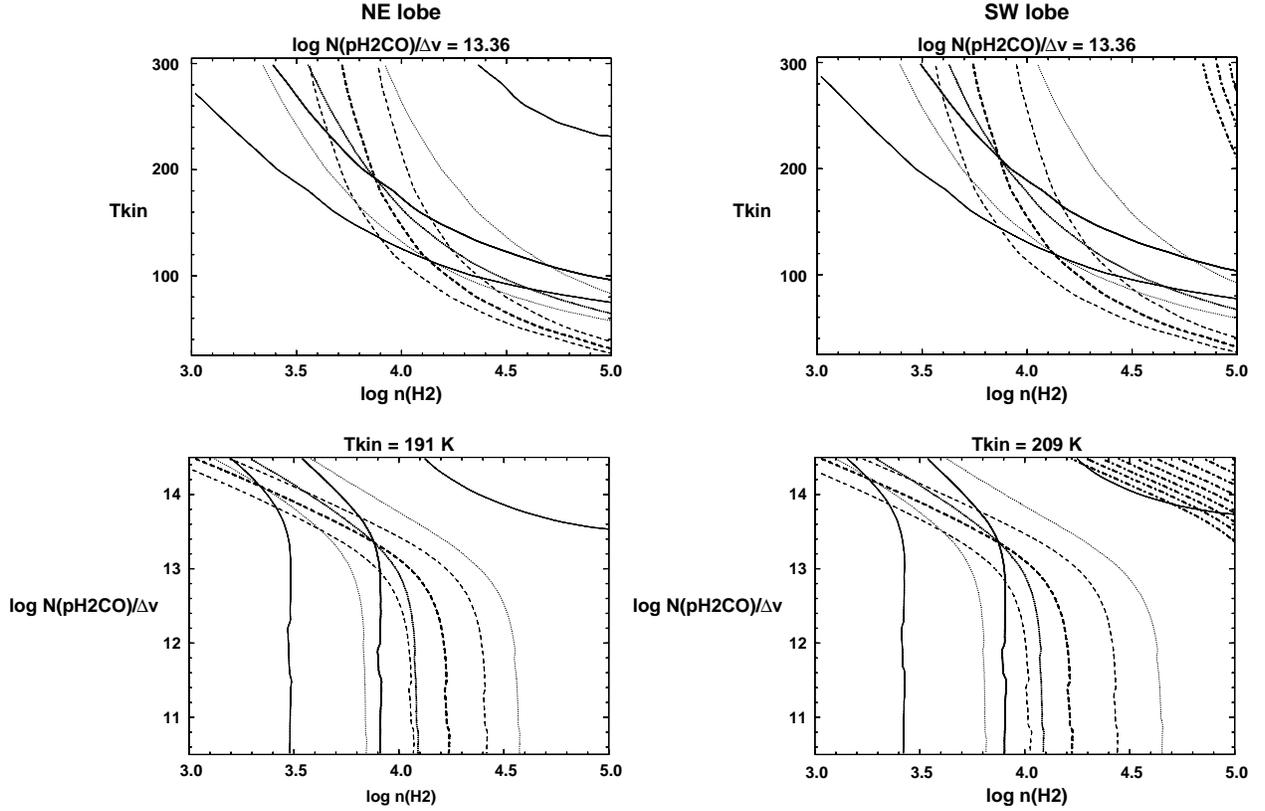}  
\caption{Results of our LVG analysis for the NE and SW lobes shown as 
cuts through the 3-D parameter space ($T_{\rm kin}$ in K, $n_{\rm H2}$ in 
cm$^{-3}$, $N_{\rm pH2CO}/\Delta v$ in cm$^{-2}$\,km$^{-1}$\,s) along a plane 
of constant para-\htco\ column 
density per velocity interval (top) and of constant kinetic temperature 
(bottom). The thick lines represent the following line ratios: 
\htco($3_{03} \to 2_{02}$)/\htco($3_{21} \to 2_{20}$) (solid), 
\htco($2_{02} \to 1_{01}$)/\htco($3_{03} \to 2_{02}$) (dashed), 
\htco($2_{02} \to 1_{01}$)/\htco($3_{21} \to 2_{20}$) (dotted). The thin lines 
outline the corresponding uncertainties given in Table~\ref{ratios}. The 
dashed-dotted line in the plots for the SW lobe denotes 
the area that is ruled out by the lower limit to the line ratio of 
\htco($2_{02} \to 1_{01}$)/\htco($5_{24} \to 4_{23}$). The adopted source size 
is $\theta_s=7\farcs5$ and the assumed para-\htco\ abundance per velocity 
gradient is $\Lambda = 1 \times 10^{-9}\,{\rm km}^{-1}\,{\rm s\,pc}$.
\label{LVGplots}
}
\end{figure}

Previous studies of \htco\ emission in starburst galaxies suggest total \htco\ 
abundances of the order of $\sim10^{-9}$ to $\sim10^{-8}$ 
\citep{huette97,wang04,martin06b}. In the high-temperature limit, the 
ortho-to-para-\htco\ ratio is $o/p=3$. However, in starburst galaxies, $o/p$ 
may be closer to 1 or 2 \citep[e.g.][]{huette97}.   
Adopting a para-\htco\ abundance per velocity gradient of
$$\Lambda=\frac{X_{\rm pH2CO}}{{\rm grad}(v)}=\frac{N_{\rm pH2CO}/\Delta v}{n_{\rm H2}} = 1 \times 10^{-9}\,{\rm km}^{-1}\,{\rm s\,pc,}$$ we derive the 
physical conditions of the molecular gas in the lobes by comparing the observed
line ratios (Table~\ref{ratios}, $\theta_s = 7\farcs5$) with those calculated 
in the model parameter space. While inter-ladder line ratios of formaldehyde 
are generally good tracers of the kinetic temperature 
and intra-ladder ratios sensitively trace the density at a given temperature 
(Section~\ref{h2co}), we find that at moderate gas densities of 
$\sim 10^4$\,cm$^{-3}$ and high kinetic temperatures of $> 100$\,K, the 
individual line ratios are not completely independent of the other model 
parameters and that a combination of the line ratios constrains the physical 
properties of the molecular gas even more tightly.
Figure~\ref{LVGplots} shows the cuts through the parameter space where the 
different derived line ratios intersect. Also plotted are the uncertainties 
given in Table~\ref{ratios}. We derive similar physical conditions for the two 
lobes, in particular a high kinetic temperature of $\sim 200$\,K and a moderate
gas density of 
$n_{\rm H2} \sim 7.4 \times 10^3$\,cm$^{-3}$ (Table~\ref{finalresults}).
Under these conditions, the 146\,GHz line is optically thick, while the optical
depth of the 218\,GHz lines is of the order of unity and the emission at 
364\,GHz is optically thin. 

If the line width $v_{\rm cloud}$ of an individual cloud is dominated by its 
velocity gradient 
($\Delta v_{\rm cloud} \approx 2r\,{\rm grad}(v)$) and the clouds are close to 
virial equilibrium, a cloud gas density of 
$n_{\rm H2} = 7.4 \times 10^3$\,cm$^{-3}$ suggests a velocity gradient of 
${\rm grad}(v) \approx 1.0$\,\kms\,pc$^{-1}$, which is in very good agreement 
with the velocity gradient deduced from high-resolution CO observations 
\citep{weiss01b}. In this case, the para-formaldehyde abundance is 
$X_{\rm pH2CO} \approx 1 \times 10^{-9}$ and the total formaldehyde abundance
$X_{\rm H2CO} \approx 2$ to $4 \times 10^{-9}$, depending on the 
ortho-to-para formaldehyde ratio.  
The mass of the molecular clouds within the beam of the telescope is  
$$ \frac{M_{\rm mol}}{M_{\odot}} = 6.828 \times 10^{-19}\,\frac{N_{\rm H2,beam}}{\rm cm^{-2}}\,\left(\frac{D}{\rm Mpc}\right)^2\,\left(\frac{\theta_{\rm beam}}{\rm arcsec}\right)^2,$$ 
where $N_{\rm H2,beam} = (N_{\rm pH2CO}/\Delta v)\,X_{\rm pH2CO}^{-1}\,\Delta v_{\rm line}\,F$ is the beam-averaged H$_2$ column density. This equation 
includes a factor 1.6 to account for dust and other molecular and atomic 
species like helium. At a distance 
of $D = 3.9$\,Mpc \citep{sakai99}, $ M_{\rm mol} = 1.7 \times 10^8\,M_{\odot}$
in the NE lobe and $ M_{\rm mol} = 1.4 \times 10^8\,M_{\odot}$ in the SW lobe.

The small-scale dilution factor $F_{sc} = F\,f_b^{-1}$ can be derived from our 
LVG calculations. Expressed in terms of cloud properties, it is 
$$ F_{sc} = f_a\,f_v = \frac{C\,r^2\,\Delta v_{\rm cloud}}{R_s^2\,\Delta v_{\rm line}}.$$ 
Assuming again that the line width of 
an individual cloud is dominated by its velocity gradient, 
$$ C\,r^3 = \frac{R_s^2\,\Delta v_{\rm line}\,F_{sc}}{2\,{\rm grad}(v)},$$
yielding 
$r = 34\,C^{-1/3}\,{\rm pc}$ in the NE lobe and $r = 33\,C^{-1/3}\,{\rm pc}$ 
in the SW lobe for a velocity gradient of 
1\,\kms\,pc$^{-1}$. 
For comparison, the maximum radius of a cloud is given by the observed line 
width and the velocity gradient  
$r_{\rm max} = \Delta v_{\rm line}/(2\,{\rm grad}(v)) = 66$\,pc and  
56\,pc in the NE and SW lobes, respectively. 
The fraction of the volume that the molecular clouds occupy within the source 
is
$$ \Phi = \frac{C\,(4/3)\,\pi\,r^3}{\pi\,R_s^2\,l} = \frac{2\,\Delta v_{\rm line}\,F_{sc}}{3\,{\rm grad}(v)\,l},$$
where $l \approx 350$\,pc is the adopted line of sight extent of the emitting 
region. For a gradient of 1\,\kms\,pc$^{-1}$, $\Phi \approx (0.30)^3$ 
or about 
30\% in each dimension of the data cube in both lobes. Thus, the observed 
volume could contain a small number of large clouds, whose diameters are 
comparable to the 
source size, or a large number of small clouds, but the presence of only a 
single cloud within the beam is ruled out.\\

The results for other choices of $\Lambda$ can be found in 
Tables~\ref{NEresults} and \ref{SWresults}. In the optically thin limit (see 
e.g.\ models N1 and N2), the small-scale dilution factor reaches its maximum 
value of $F_{sc} = 1$ and the temperature-density plots are almost independent 
of the assumed \htco\ column density per velocity interval. Thus, there is a 
lower limit to the kinetic temperatures and an upper limit to the gas densities
consistent with our observations. We do not find a solution (corresponding to 
the intersection of the curves for  
the line ratios) in the calculated parameter space for 
$T_{\rm kin} \lesssim 140$\,K (NE lobe) and $T_{\rm kin} \lesssim 155$\,K (SW
lobe), respectively. The upper limit to the gas density is 
$n_{\rm H2} \sim 2.5 \times 10^4$ in both lobes.\\

The assumed source size $\theta_s$ also has a significant influence on the 
derived parameters (Table~\ref{finalresults}). A larger source size results in 
a lower kinetic temperature, a higher H$_2$ density and a smaller small-scale 
dilution factor. But even if the \htco\ emitting region is as large as 
10\arcsec, the kinetic temperature in the lobes is $\sim 170$\,K, which is much
higher than the dust temperature of $T_{\rm dust} = 48$\,K \citep{colbert99} or
the commonly assumed kinetic temperatures of 30\,K to 100\,K.
To summarize, the dominant component of the molecular gas in the
lobes of M\,82 seems to be in a warm phase of moderate density.\\

\subsection{Comparison with Other Excitation Studies}
\label{warm}

Numerous previous studies have used molecular lines to derive the 
physical properties of the molecular gas in M\,82. Frequently, the small number
of lines available to constrain the free parameters has led to the adoption 
of a kinetic temperature of $T_{\rm kin} \sim 50$\,K, similar to the 
dust temperature $T_{\rm dust} = 48$\,K \citep{colbert99}. This assumption is 
supported by the results of early analyzes of the low-$J$ CO lines 
\citep[e.g.][]{wild92} and, more recently, by the ammonia study of 
\citet[][see below]{weiss01a}. 

The most comprehensive data base exists for CO and $^{13}$CO lines, which have 
been searched for up to $J = 7 \to 6 $ \citep{mao00} and  $J = 6\to 5 $ 
\citep{ward03}, respectively. Based on this large collection of CO lines, 
\citet{mao00} confirm that a one-component LVG model can not fit all the 
observed line intensities simultaneously. Thus, in order to derive the 
properties of the high-excitation component, they restrict their LVG analysis 
to the high-J excitation lines in the submm-range, i.e. CO($7 \to 6$), 
CO($4 \to 3$), CO($3 \to 2$), and $^{13}$CO($3 \to 2$).
In the most recent multi-line CO study, \cite{ward03} analyze the CO and 
$^{13}$CO emission at all available transitions using a two-component 
plane-parallel LVG model. The physical properties of the high-excitation 
molecular gas phase derived from these comprehensive excitation 
studies are compared to our results in Table~\ref{otherlvg}. Considering that
our investigation is completely independent of the other studies, the results
are remarkably similar. The physical properties derived from our \htco\ data 
and our LVG model adopting a spherically symmetric cloud morphology lie well 
within the range of values found for the high-excitation component of the 
two-component LVG model by \citet{ward03}. In fact, our kinetic temperature 
is only 20\,K to 40\,K higher than their median value, which is influenced by 
the limited temperature range considered, and their median density is lower by 
only a factor of $\sim 2$ if the effect of using a 
plane-parallel model versus a spherically symmetric one is taken into account.
The most striking difference between our results and the values derived by 
\citet{mao00} is their lower kinetic temperature. 
A possible explanation for this discrepancy is a small contribution of the 
cooler low-excitation component to the submm CO line emission that would lower 
the average kinetic temperature of the one-component model. 

The presence of a warm gas component in nearby starburst 
galaxies with $T_{\rm kin} = 50$ to 440\,K was also inferred from the 
detection of highly excited NH$_3$ lines \citep{mauers03}. In this study, M\,82
was the notable exception in not showing evidence for a warm molecular gas 
component in its NH$_3$ emission. The ammonia inversion lines detected towards 
M\,82 suggest a rotational temperature of only 29\,K and thus a kinetic 
temperature of only $\sim60$\,K in the SW lobe of M\,82 \citep{weiss01a}. 
In contrast, \citet{rigopoulou02} find a warm molecular
gas component with $T_{\rm kin} \sim 150$\,K in a sample of starburst and 
Seyfert galaxies including M\,82 by analyzing IR rotational H$_2$ emission 
lines.
Possible mechanisms that could heat a large fraction of the molecular 
gas to temperatures of $\sim 150$\,K include shocks, strong UV and/or X-ray 
irradiation (PDR, XDR) \citep[e.g.][]{rigopoulou02, fuente05} as well as the 
more uniform cosmic ray heating \citep{bradford03}, in short processes that are
likely to be found in an active environment such as a starburst.

A comparison of the \htco\ line widths and the derived molecular gas mass with 
the values derived from the CO studies suggests 
that the \htco\ lines trace approximately the same gas as traced by the 
ubiquitous CO lines (Table~\ref{otherlvg}). According to our data, each lobe 
contains warm molecular gas of about 
$M_{\rm mol} \sim 1.5 \times 10^8\,M_{\odot}$, which is in good agreement with 
the estimate by \citet{mao00} of $M_{\rm mol} \sim 1-10 \times 10^8\,M_{\odot}$
and the mass of the molecular ring of  
$M_{\rm mol} \sim 2.0 \times 10^8\,M_{\odot}$ derived by \citet{ward03} from
the integrated intensity of their CO($6 \to 5$) map. Having modeled the 
physical conditions and the conversion factor $X_{\rm CO}$ at 18 positions 
along the circumnuclear ring as traced by high-resolution low-J CO maps, 
\citet{weiss01b} find a molecular gas mass of 
$M_{\rm mol} \sim 2.5 \times 10^8\,M_{\odot}$ in the ring, whereas applying 
the standard conversion factor 
$X_{\rm CO}=1.6 \times 10^{20}$\,cm$^{-2}$\,(K\,km\,s$^{-1}$)$^{-1}$ to their 
data leads to a total mass of $M_{\rm mol} \sim 7.1 \times 10^8\,M_{\odot}$.

In contrast to the results of the CO and \htco\ investigations, the observed 
ammonia inversion lines seem to arise from cold, probably well shielded 
regions within the ISM \citep{mauers03}. The physical properties derived 
in our analysis and the CO studies suggest that the majority of the gas is in 
a warm state of comparatively low density 
($n_{\rm H2} \sim 7 \times 10^3\,{\rm cm^{-3}}$)), which fits well with the 
idea that M\,82 is in an advanced phase of a starburst. In such an 
environment, ammonia is easily photodissociated outside of the few remaining 
well-shielded regions \citep{fuente93}, which may
explain the extremely low NH$_3$ column densities reported by \citet{weiss01a}.

\section{METHANOL EMISSION FROM THE MOLECULAR LOBES OF M\,82}
\label{methanol}

\begin{figure}
\plotone{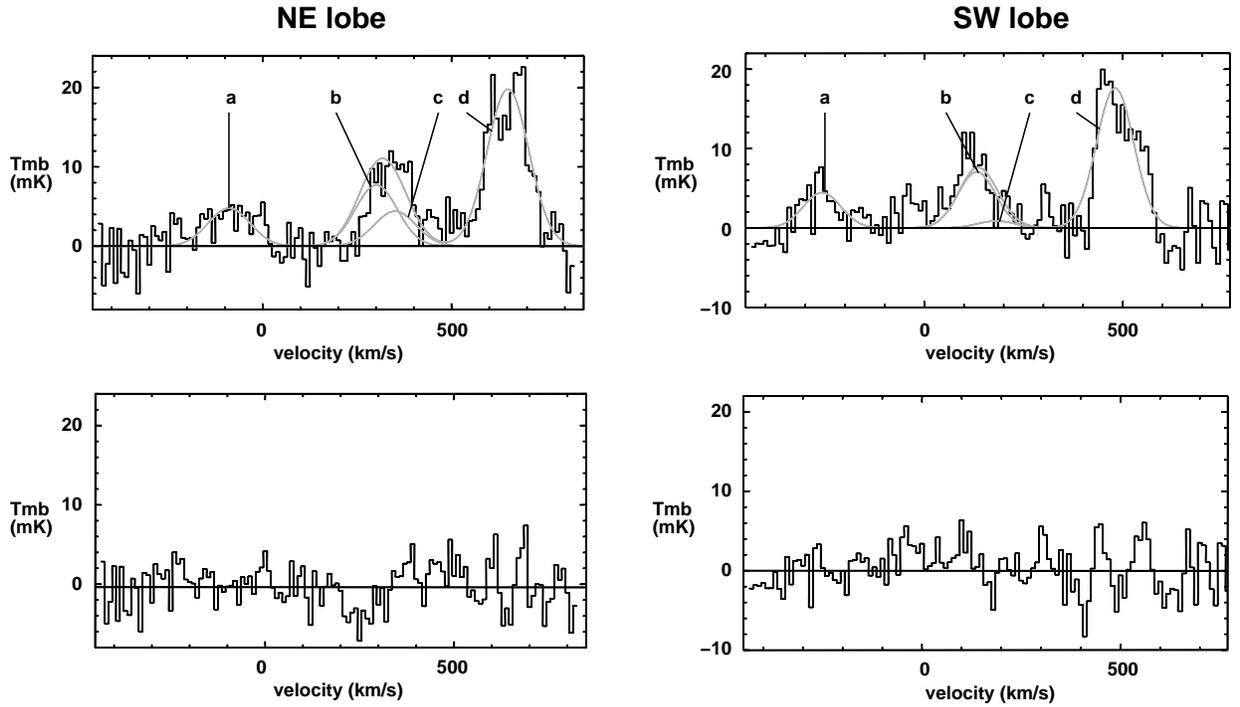}  
\caption{{\bf Top:} Gaussian fits derived from the LVG analysis 
(Section~\ref{methanol}) superposed on the 218\,GHz spectra of the NE lobe 
(left) and the SW lobe (right); the curves show the Gaussian fit to each 
individual line as well as the spectrum resulting from a superposition of all 
identified lines: a = \htco($3_{21} \to 2_{20}$), 
b = \htco($3_{22} \to 2_{21}$), c = CH$_3$OH($4_2 \to 3_1\, E$), 
d = \htco($3_{03} \to 2_{02}$).  
{\bf Bottom:} Residual spectra after subtraction of the Gaussian fits.
\label{SWlvg}
}
   \end{figure}

The detection of the \htco($3_{21} \to 2_{20}$) line in both molecular lobes 
strongly suggests that there is \htco($3_{22} \to 2_{21}$) emission at these 
positions too. On the other hand, \citet{martin06a} very recently reported 
the detection of methanol emission from the lobes of M\,82 including the 
highly excited $5_k \to 4_k$ transition. Thus, it is reasonable to expect 
CH$_3$OH($4_2 \to 3_1\, E$) emission to originate from the two lobes as well. 
As a result, the observed spectral features at 218.5\,GHz in both lobes most 
likely consist of the strongly blended emission of the 
\htco($3_{22} \to 2_{21}$)  and the CH$_3$OH($4_2 \to 3_1\, E$) lines at 
218.48\,GHz  and at 218.44\,GHz, respectively. Because of this blending, the 
spectral feature at $\sim 218.5$\,GHz was ignored in the LVG analysis in 
Section~\ref{lvg}. Now we can use the results of our LVG models to derive the 
expected strengths of the \htco($3_{22} \to 2_{21}$) emission for different 
physical conditions (Tables~\ref{NEresults} and~\ref{SWresults}). 
Interestingly, the computed values for each lobe vary by no more than 5\% 
throughout the range of model parameters considered, which indicates that the 
\htco($3_{22} \to 2_{21}$) emission does not add significant constraints to 
the LVG analysis, probably because the similar \htco($3_{21} \to 2_{20}$) line 
is already included. In order to deconvolve the highly blended lines, we fixed 
the integrated intensity of the \htco($3_{22} \to 2_{21}$) emission to the 
value derived from the models and fitted the five lines with the usual 
constraints (Table~\ref{lines}). Fig.~\ref{SWlvg} shows the resulting Gaussian 
fits and the residual spectra in both lobes. The derived integrated 
intensities are summarized in Table~\ref{lvgfits}. 

In the SW lobe, the \htco($3_{22} \to 2_{21}$) emission predicted by the LVG 
analysis agrees within $3\sigma$ with the previously derived integrated 
intensity, and the estimated contribution of the methanol line is well below 
the  uncertainty of the mean of the fitted \htco\ line. In the NE lobe, the 
methanol line appears to be about six times stronger, while the predicted 
\htco($3_{22} \to 2_{21}$) emission is comparable to that of the SW lobe. 
The noise level in the NE residual spectrum is slightly increased at the 
position of the subtracted 
blended lines. However, this is also the case at the position of the 
subtracted \htco($3_{03} \to 2_{02}$) line and may be due to a 
deviation of the line profile from a Gaussian shape.
The result that the CH$_3$OH($4_2 \to 3_1\, E$) emission of the NE lobe is much
stronger than that of the SW lobe agrees very well with the line strengths 
reported by \citet{martin06a}. Their intensities of the blended methanol lines 
at 97\,GHz and 157\,GHz are similar in both lobes. However, the 
CH$_3$OH($5_k \to 4_k$) transitions at 242\,GHz, which require high excitation 
levels similar to those needed to observe the CH$_3$OH($4_2 \to 3_1\, E$) 
transition, are stronger in the NE lobe than in the SW lobe, which might 
indicates a difference in excitation.

\section{CONCLUSIONS}

\begin{enumerate}
\item We have detected several para-\htco\ emission lines at 146\,GHz and 
218\,GHz in both molecular lobes near the center of the nearby starburst galaxy
M\,82. To our knowledge, our study constitutes the first dedicated search for  
\htco\ transitions with $K_a > 1$ and the first detection of 
\htco($3_{22} \to 2_{21}$) emission outside of the Galactic neighborhood.
\item Our line ratio analysis using an LVG model with spherical cloud 
symmetry suggests
the presence of warm ($T_{\rm kin} \sim 200$\,K), moderately dense 
($n_{\rm H2} \sim 7 \times 10^3\,{\rm cm^{-3}}$) molecular gas in the lobes 
near the center 
of the starburst activity. The ratio of column density to volume 
density and the small small-scale filling factor may indicate that the 
observed molecular gas forms large complexes of comparatively low-density 
molecular gas, 
possibly extended envelopes around a few remaining well-shielded cores.  
\item The physical properties of the molecular gas derived from our \htco\ 
data are in very good agreement with parameters of the high-excitation 
component in the lobes of M\,82 found in recent comprehensive CO studies, but 
differ from the results of a multi-line NH$_3$ investigation. Thus, the
para-\htco\ lines seem to trace the probably dominant high-excitation 
component of the molecular gas in M\,82 very well, while the NH$_3$ emission 
might originate predominantly from cold cloud cores. 
\item The total mass of the molecular gas observed in the lobes is a few times 
$10^8\,M_{\sun}$, similar to the mass determinations of other molecular line 
studies.
\item The selected para-\htco\ transitions are good tracers of the 
molecular gas in starburst galaxies, even if a large fraction of the 
gas is in a warm phase ($T_{\rm kin} \sim 150$\,K) of moderate density 
($n_{\rm H2} \lesssim 10^4\,{\rm cm^{-3}}$).
\item  We find strong evidence for CH$_3$OH($4_2 \to 3_1\, E$) emission in the
spectrum of the NE lobe at 218\,GHz and also estimate the intensity of this 
emission line in the SW lobe. We thus confirm the detection of methanol in 
M\,82 by \citet{martin06b} with measurements of another transition line that 
are consistent with the earlier findings.
\end{enumerate}

\acknowledgments

 We wish to thank the staff at Pico Veleta/Granada and at the JCMT/MKSS/JAC 
for their support during the observations. We also thank the anonymous referee
whose thoughtful comments greatly helped to improve the manuscript. S.M.\ 
acknowledges a travel grant 
from the Natural Science and Engineering Research Council of Canada (NSERC) 
administered by the National Research Council of Canada (NRC) for the 
observations at the JCMT. E.R.S. acknowledges a Discovery Grant from NSERC.  
This work has made use of the following software and resources: GILDAS, specx, 
Statistiklabor (FU Berlin, CeDiS), asyerr (Seaquist \& Yao), 
the JPL Catalog of spectral lines, The Cologne Database for Molecular 
Spectroscopy \citep[CDMS][]{mueller05} and NASA's Astrophysics Data System 
Bibliographic Services (ADS). 

{\it Facilities:} \facility{IRAM:30m (CD,HERA)}, \facility{JCMT (B3)}.

\clearpage

\begin{deluxetable}{lccc}
\tabletypesize{\small}
\tablewidth{0pt}
\tablecaption{Summary of the observed para- ($p$) and ortho-\htco\ ($o$) 
              transitions. \label{obs}}
\tablehead{
\colhead{Transition} &  \colhead{$\nu$/[GHz]} & \colhead{$\theta_b$\tablenotemark{a}/[\arcsec]} 
& \colhead{rms\tablenotemark{b}/[mK]}
}
\startdata
\htco($2_{02} \to 1_{01}$) $\quad p$ & 145.6029 & 17 & 1.9, 1.8\\
\htco($3_{03} \to 2_{02}$) $\quad p$ & 218.2222 & 11 & 2.7, 2.4\\
\htco($3_{22} \to 2_{21}$) $\quad p$ & 218.4756 & 11 & 2.7, 2.4\\
\htco($3_{21} \to 2_{20}$) $\quad p$ & 218.7601 & 11 & 2.7, 2.4\\
\htco($5_{24} \to 4_{23}$) $\quad p$ & 363.9459 & 13 & 5.4, n/a\\
\htco($5_{42} \to 4_{41}$) $\quad p$ & 364.1032 & 13 & 5.4, n/a\\
\htco($5_{41} \to 4_{40}$) $\quad p$ & 364.1032 & 13 & 5.4, n/a\\
                           &        &    & \\
\htco($5_{33} \to 4_{32}$) $\quad o$ & 364.2751 & 13 & 5.4, n/a\\
\htco($5_{32} \to 4_{31}$) $\quad o$ & 364.2889 & 13 & 5.4, n/a\\
\enddata
\tablenotetext{a}{half power beam width (FWHM)}
\tablenotetext{b}{rms in $T_{\rm mb}$ at 10\,\kms\ resolution at the positions of the SW lobe and the NE lobe, respectively}
\end{deluxetable}

\clearpage

\begin{deluxetable}{lcccc} 
\tabletypesize{\footnotesize}
\tablewidth{0pt}
\tablecaption{Input parameters and results of the Gaussian fits to the observed emission lines. \label{lines}}
\tablehead{
\colhead{Parameter} & \colhead{NE lobe} & \colhead{SW lobe} & 
\colhead{SW lobe/narrow} & \colhead{SW lobe/wide} 
}
\startdata
$v_0$ \,[\kms]       & $301.0\pm3.6$ & $132.2\pm4.8$ & $106.3\pm5.8$ & $174.3\pm12.7$ \\
$\Delta v_{\rm line}$ \,[\kms]  & $130.7\pm4.1$ & $111.2\pm6.4$ &  $48.3\pm10.5$ &  $77.1\pm18.6$ \\                      
\tableline 
$\int{T_{\rm mb}\,dv}$(\htco($2_{02} \to 1_{01}$))   & $3.69\pm0.10$                 & $2.76\pm0.10$          & $1.34\pm0.07$ & $1.60\pm0.15$  \\
$\int{T_{\rm mb}\,dv}$(HC$_3$N($16 \to 15$))         & $0.46\pm0.11$                 & $0.70\pm0.09$          & $<0.19$     & $0.42\pm0.08$  \\
\tableline 
$\int{T_{\rm mb}\,dv}$(\htco($3_{03} \to 2_{02}$))   & $2.75\pm0.18$ (2.73)          & $2.09\pm0.16$ (2.09)   & $0.98\pm0.08$ & $1.10\pm0.12$ \\ 
$\int{T_{\rm mb}\,dv}$(\htco($3_{22} \to 2_{21}$))   & {\it 0.05$\pm$0.17 (0.11)}  & {\it 1.09$\pm$0.15 (1.10)} & {\it 0.57$\pm$0.08} & {\it 0.41$\pm$0.10} \\ 
$\int{T_{\rm mb}\,dv}$(\htco($3_{21} \to 2_{20}$))  & $0.67\pm0.15$ ($0.68$)        & $0.53\pm0.14$ (0.53)   & $0.25\pm0.09$ & $0.21\pm0.08$ \\
$\int{T_{\rm mb}\,dv}$(CH$_3$OH($4_2 \to 3_1\, E$)) & {\it 1.45$\pm$0.21 (1.39)}    & {\it ($<$ 0.27)}       &              &              \\
$\int{T_{\rm mb}\,dv}$(HC$_3$N($24 \to 23$))         &               (0.17)          & ($< 0.27$)                 &              &              \\
\tableline 
$\int{T_{\rm mb}\,dv}$(\htco($5_{24} \to 4_{23}$))   &                               & $< 0.54$               &              &              \\ 
$\int{T_{\rm mb}\,dv}$(\htco($5_{4x} \to 4_{4y}$))\tablenotemark{a}   &                               & $< 0.54$               &              &              \\ 
$\int{T_{\rm mb}\,dv}$(\htco($5_{3x} \to 4_{3y}$))\tablenotemark{b}   &                               & $< 0.54$               &              &              \\  
\enddata 
\tablenotetext{a}{integrated intensity of the highly blended lines \htco($5_{42} \to 4_{41}$) and \htco($5_{41} \to 4_{40}$)}
\tablenotetext{b}{integrated intensity of the highly blended lines \htco($5_{33} \to 4_{32}$) and \htco($5_{32} \to 4_{31}$)}
\tablecomments{All integrated intensities in K\,\kms\ (main-beam temperature 
scale) and before correction for
beam width. The last two columns show the line intensities in the SW lobe that
result from the assumption of two velocity components (``narrow'' and 
``wide''). At each position, the velocity $v_0$, the line width 
$\Delta v_{\rm line}$ and their errors were derived by simultaneously fitting 
Gaussians with a fixed frequency offset and the same (unknown) line width to 
the detected lines in the 218\,GHz spectrum (NE lobe: 4 components, SW lobe: 3 
components, SW lobe/narrow and wide: 2 components to the 
\htco($3_{03} \to 2_{02}$) line alone). The derived velocities and line widths 
were then fixed for the fits to the other spectra.
Values in parentheses show the results 
of a simultaneous fit to all five potentially present lines in the 218\,GHz 
spectra 
(\htco($3_{03} \to 2_{02}$), HC$_3$N($24 \to 23$), CH$_3$OH($4_2 \to 3_1\, E$),
 \htco($3_{22} \to 2_{21}$), and \htco($3_{21} \to 2_{20}$)); values in italics
indicate 
highly blended lines, whose fit is uncertain; upper limits in this table 
are 3$\sigma$ limits to the integrated intensities of the spectral lines using 
the adopted line widths.}
\end{deluxetable}

\clearpage

\begin{deluxetable}{lccc|ccc}
\tabletypesize{\footnotesize}
\tablewidth{0pt}
\tablecaption{Beam-corrected line ratios derived from the observed line intensities and the assumed source size $\theta_s$ \label{ratios}}
\tablehead{
\colhead{line intensity ratio / source, $\theta_s$\tablenotemark{a}} & \colhead{NE, 5\arcsec} & \colhead{NE, 7\farcs5} & \colhead{NE, 10\arcsec} & \colhead{SW, 5\arcsec} & \colhead{SW, 7\farcs5} & \colhead{SW, 10\arcsec}}  
\startdata 
\htco($2_{02} \to 1_{01}$)/\htco($3_{03} \to 2_{02}$) & $ 2.88_{-0.22}^{+0.20}$ & $ 2.61_{-0.20}^{+0.18}$ &  $2.36_{-0.18}^{+0.16}$ & $ 2.84_{-0.26}^{+0.23}$ & $ 2.58_{-0.24}^{+0.21}$ &  $2.33_{-0.21}^{+0.19}$ \\
\htco($2_{02} \to 1_{01}$)/\htco($3_{21} \to 2_{20}$) & $11.91_{-3.11}^{+2.21}$ & $10.79_{-2.81}^{+1.99}$ &  $9.75_{-2.54}^{+1.80}$ & $11.26_{-3.31}^{+2.31}$ & $10.20_{-3.00}^{+2.09}$ &  $9.22_{-2.71}^{+1.89}$ \\
\htco($3_{03} \to 2_{02}$)/\htco($3_{21} \to 2_{20}$) & $ 4.13_{-1.12}^{+0.80}$ & $ 4.13_{-1.12}^{+0.80}$ &  $4.13_{-1.12}^{+0.80}$ & $ 3.96_{-1.21}^{+0.85}$ & $ 3.96_{-1.21}^{+0.85}$ &  $3.96_{-1.21}^{+0.85}$ \\
\htco($2_{02} \to 1_{01}$)/\htco($5_{24} \to 4_{23}$) & & & & $> 8.28$ & $> 7.84$ & $>7.40$ \\
\enddata 
\tablenotetext{a}{selected lobe, assumed source size $\theta_s$}
\end{deluxetable}

\clearpage

\begin{deluxetable}{lccc|ccc}
\tabletypesize{\footnotesize}
\tablewidth{0pt}
\tablecaption{Results of our LVG calculations and the derived physical properties assuming $\Lambda = 1 \times 10^{-9}$\,km$^{-1}$\,s\,pc. \label{finalresults}}
\tablehead{
\colhead{parameter / source, $\theta_s$\tablenotemark{a}} & \colhead{NE, 5\arcsec} & \colhead{NE, 7\farcs5} & \colhead{NE, 10\arcsec} & \colhead{SW, 5\arcsec} & \colhead{SW, 7\farcs5} & \colhead{SW, 10\arcsec}}
\startdata
$T_{\rm kin}$ [K]                &   230 &   191 &   161 &   256 &   209 &   181\\
$\log{n_{\rm H2}}$ [cm$^{-3}$]   &  3.67 &  3.87 &  4.04 &  3.67 &  3.87 &  4.02\\
$\log{N_{\rm pH2CO}/\Delta v}$\tablenotemark{b} [cm$^{-2}$\,km$^{-1}$\,s] & 13.16 & 13.36 & 13.53 & 13.16 & 13.36 & 13.51\\
$\tau_{145.6}$\tablenotemark{c}    & 2.508 & 3.628 & 4.306 & 2.473 & 3.551 & 4.173\\
$\tau_{218.2}$\tablenotemark{c}  & 0.636 & 1.345 & 1.933 & 0.642 & 1.355 & 1.943\\
$\tau_{218.5}$\tablenotemark{c}  & 0.261 & 0.432 & 0.520 & 0.274 & 0.455 & 0.557\\
$\tau_{218.8}$\tablenotemark{c}  & 0.153 & 0.273 & 0.343 & 0.163 & 0.291 & 0.368\\
$\tau_{364.0}$\tablenotemark{c}  & 0.003 & 0.007 & 0.010 & 0.004 & 0.008 & 0.011\\
$F$\tablenotemark{c,d}           & 0.025 & 0.019 & 0.013 & 0.037 & 0.018 & 0.013\\
$F_{sc}$\tablenotemark{c,e}      & 0.24  & 0.11  & 0.08  & 0.23  & 0.11  & 0.08 \\
\tableline
$p = n_{\rm H2}\, T_{\rm kin}$       [$10^6\,{\rm cm}^{-3}$\,K]  & 1.1 & 1.4 & 1.8 & 1.2 & 1.6 & 1.9 \\
$N_{\rm pH2CO,beam}$\tablenotemark{f}  [$10^{13}\,{\rm cm}^{-2}$] & 4.7 & 5.7 & 5.8 & 6.0 & 4.6 & 4.7 \\
$N_{\rm H2,beam}$\tablenotemark{g}    [$10^{22}\,{\rm cm}^{-2}$] & 4.7 & 5.7 & 5.8 & 6.0 & 4.6 & 4.7 \\
$M_{\rm mol}$\tablenotemark{h}            [$10^8\,M_{\odot}$]    & 1.4 & 1.7 & 1.7 & 1.8 & 1.4 & 1.4 \\
\enddata
\tablenotetext{a}{selected lobe, assumed source size $\theta_s$}
\tablenotetext{b}{calculated from $\Lambda$ and $\log{n_{\rm H2}}$}
\tablenotetext{c}{evaluated at the nearest grid point}
\tablenotetext{d}{dilution factor $F = T_{\rm obs}/T_{\rm lvg}$ in a 
17\arcsec\ beam; the 218\,GHz data are scaled to a 17\arcsec\ beam assuming a 
source size of $\theta_s$}
\tablenotetext{e}{small-scale dilution factor $F_{sc} = F\,f_b^{-1} = f_a\,f_v = F/0.163$ for $\theta_s=$7\farcs5.}
\tablenotetext{f}{beam-averaged column density $N_{\rm pH2CO,beam} = (N_{\rm pH2CO}/\Delta v)\,\Delta v_{\rm line}\,F$}
\tablenotetext{g}{beam-averaged column density, assuming $X_{\rm pH2CO} = 1 \times 10^{-9}$}
\tablenotetext{h}{molecular gas mass, assuming $X_{\rm pH2CO} = 1 \times 10^{-9}$ and $D = 3.9$\,Mpc}
\end{deluxetable}

\clearpage

\begin{deluxetable}{lcccccccccccc}
\tabletypesize{\footnotesize}
\tablewidth{0pt}
\rotate
\tablecaption{Results of our LVG calculations for various physical conditions 
of the molecular gas in the NE lobe. \label{NEresults}}
\tablehead{
\colhead{model} & \colhead{$T_{\rm kin}$} & \colhead{$\log{n_{\rm H2}}$} & \colhead{$\log{N_{\rm pH2CO}/\Delta v}$} & \colhead{$X_{\rm pH2CO}/{\rm grad}\,v$}  & \colhead{$\tau_{146}$\tablenotemark{a}} & \colhead{$\tau_{218.2}$\tablenotemark{a}} & \colhead{$\tau_{218.5}$\tablenotemark{a}} & \colhead{$\tau_{218.8}$\tablenotemark{a}} & \colhead{$\tau_{364}$\tablenotemark{a}} & \colhead{$F$\tablenotemark{a,b}} & \colhead{$F_{sc}$\tablenotemark{a,c}} & \colhead{$\int{T_{\rm mb}\,dv}_{218.5}$\tablenotemark{a,d}}\\
      &  [K]      &  [cm$^{-3}$]   & [cm$^{-2}$\,km$^{-1}$\,s] & [$10^{-9}$\,km$^{-1}$\,s\,pc] & & & & & & & & [K\,\kms] 
}  
\startdata 
 T1 &  150 & 4.24 & 12.73 &  0.100  & 0.78  & 0.22 & 0.09 & 0.06 & 0.002 & 0.059 & 0.36 & 0.53 \\
 T2 &  175 & 4.02 & 13.18 &  0.469  & 2.41  & 0.79 & 0.29 & 0.18 & 0.006 & 0.027 & 0.16 & 0.54 \\
 T3 &  200 & 3.79 & 13.47 &  1.55   & 4.72  & 1.81 & 0.53 & 0.33 & 0.008 & 0.020 & 0.12 & 0.55 \\
 T4 &  225 & 3.64 & 13.61 &  3.03   & 6.24  & 2.23 & 0.58 & 0.34 & 0.007 & 0.023 & 0.14 & 0.55 \\
 T5 &  250 & 3.52 & 13.75 &  5.51   & 9.52  & 4.02 & 0.84 & 0.51 & 0.010 & 0.017 & 0.11 & 0.54 \\
T6  &  275 & 3.40 & 13.88 &  9.81   & 11.96 & 5.24 & 0.95 & 0.59 & 0.010 & 0.016 & 0.10 & 0.55 \\
T7  &  300 & 3.28 & 13.99 & 16.65   & 15.07 & 6.82 & 1.11 & 0.67 & 0.011 & 0.016 & 0.10 & 0.55 \\
\tableline
 N1 &  142 & 4.38 & 12.00 &  0.0135 & 0.15  & 0.04 & 0.19 & 0.13 & 0.0007& 0.202 & 1.24 &  n/a\tablenotemark{e} \\
 N2 &  146 & 4.31 & 12.50 &  0.0503 & 0.49  & 0.14 & 0.06 & 0.04 & 0.002 & 0.078 & 0.48 & 0.52 \\
 N3 &  159 & 4.13 & 13.00 &  0.2407 & 1.54  & 0.46 & 0.19 & 0.12 & 0.004 & 0.036 & 0.22 & 0.53 \\
 N4 &  206 & 3.76 & 13.50 &  1.7842 & 4.68  & 1.81 & 0.53 & 0.33 & 0.008 & 0.020 & 0.12 & 0.55 \\
 N5 &  298 & 3.28 & 14.00 & 17.0392 & 15.04 & 6.80 & 1.11 & 0.67 & 0.011 & 0.016 & 0.10 & 0.55 \\
\enddata
\tablecomments{For the T models, a kinetic temperature was specified, while the N models used a specific para-\htco\ column density per velocity interval as an input parameter.}
\tablenotetext{a}{evaluated at the nearest grid point}
\tablenotetext{b}{dilution factor $F = T_{\rm obs}/T_{\rm lvg}$ in a 
17\arcsec\ beam; the 218\,GHz data are scaled to a 17\arcsec\ beam assuming a 
source size of $\theta_s=$7\farcs5}
\tablenotetext{c}{small-scale dilution factor $F_{sc} = F\,f_b^{-1} = f_a\,f_v = F/0.163$ for $\theta_s=$7\farcs5.}
\tablenotetext{d}{calculated integrated intensity of the 
                  \htco($3_{21} \to 2_{20}$) line} 
\tablenotetext{e}{value ambiguous}
\end{deluxetable}

\clearpage

\begin{deluxetable}{lcccccccccccc}
\tabletypesize{\footnotesize}
\tablewidth{0pt}
\rotate
\tablecaption{Results of our LVG calculations for various physical conditions of the molecular gas in the SW lobe. \label{SWresults}}
\tablehead{
\colhead{model} & \colhead{$T_{\rm kin}$} & \colhead{$\log{n_{\rm H2}}$} & \colhead{$\log{N_{\rm pH2CO}/\Delta v}$} & \colhead{$X_{\rm pH2CO}/{\rm grad}\,v$}  & \colhead{$\tau_{146}$\tablenotemark{a}} & \colhead{$\tau_{218.2}$\tablenotemark{a}} & \colhead{$\tau_{218.5}$\tablenotemark{a}} & \colhead{$\tau_{218.8}$\tablenotemark{a}} & \colhead{$\tau_{364}$\tablenotemark{a}} & \colhead{$F$\tablenotemark{a,b}} & \colhead{$F_{sc}$\tablenotemark{a,c}} & \colhead{$\int{T_{\rm mb}\,dv}_{218.5}$\tablenotemark{a,d}}\\ 
      &  [K]      &  [cm$^{-3}$]   & [cm$^{-2}$\,km$^{-1}$\,s] & [$10^{-9}$\,km$^{-1}$\,s\,pc] & & & & & & & & [K\,\kms] 
}
\startdata  
 T2 &  175 & 4.17 & 12.83 &  0.148  & 0.943 & 0.287 & 0.130 & 0.085 & 0.004  & 0.038 & 0.23 & 0.42 \\
 T3 &  200 & 3.93 & 13.28 &  0.727  & 3.014 & 1.024 & 0.366 & 0.230 & 0.006  & 0.022 & 0.13 & 0.43 \\
 T4 &  225 & 3.78 & 13.45 &  1.52   & 4.605 & 1.832 & 0.561 & 0.354 & 0.009  & 0.016 & 0.10 & 0.43 \\
 T5 &  250 & 3.65 & 13.60 &  2.89   & 5.790 & 2.393 & 0.671 & 0.423 & 0.010  & 0.015 & 0.09 & 0.43 \\
 T6 &  275 & 3.53 & 13.72 &  5.03   & 7.726 & 2.948 & 0.717 & 0.432 & 0.008  & 0.018 & 0.11 & 0.43 \\
 T7 &  300 & 3.43 & 13.86 &  8.74   & 11.78 & 5.282 & 1.022 & 0.625 & 0.011  & 0.014 & 0.08 & 0.43 \\
\tableline
 N1 &  159 & 4.34 & 12.00 &  0.0148 & 0.153 & 0.041 & 0.020 & 0.014 & 0.0006 & 0.203 & 1.25 & 0.41 \\
 N2 &  164 & 4.27 & 12.50 &  0.0551 & 0.472 & 0.140 & 0.065 & 0.044 & 0.002  & 0.064 & 0.39 & 0.42 \\
 N3 &  183 & 4.10 & 13.00 &  0.2579 & 1.492 & 0.470 & 0.201 & 0.130 & 0.005  & 0.029 & 0.18 & 0.43 \\
 N4 &  230 & 3.75 & 13.50 &  1.8258 & 4.575 & 1.827 & 0.564 & 0.358 & 0.009  & 0.016 & 0.10 & 0.43 \\
\enddata
\tablecomments{For the T models, a kinetic temperature was specified, while the N models used a specific para-\htco\ column density per velocity interval as an input parameter.}
\tablenotetext{a}{evaluated at the nearest grid point}
\tablenotetext{b}{dilution factor $F = T_{\rm obs}/T_{\rm lvg}$ in a 
17\arcsec\ beam; the 218\,GHz data are scaled to a 17\arcsec\ beam assuming a 
source size $\theta_s=$7\farcs5}
\tablenotetext{c}{small-scale dilution factor $F_{sc} = F\,f_b^{-1}=f_a\,f_v$}
\tablenotetext{d}{calculated integrated intensity of the 
                  \htco($3_{21} \to 2_{20}$) line} 
\end{deluxetable}

\clearpage

\begin{deluxetable}{lcccc}
\tabletypesize{\footnotesize}
\tablewidth{0pt}
\rotate
\tablecaption{Comparison of our results to those of other excitation studies.\label{otherlvg}}
\tablehead{\colhead{parameter/study} & \colhead{our data (NE/SW)} & \colhead{Ward/median (NE/SW)\tablenotemark{a,b}} & \colhead{Ward/range (NE/SW)\tablenotemark{a,c}} & \colhead{Mao (NE/SW)\tablenotemark{d}}} 
\startdata
$\theta_b$        [arcsec]                             & 17        & 24.4      & 24.4                    & 22\\
$\Delta v_{\rm line}$ [\kms]                           & 131/111   & 180/160   & 180/160                 & $\sim120$/$\sim130$\tablenotemark{e}  \\
$T_{\rm kin}$     [K]                                  & 191/209   & 170/170   & $>50$/$>50$             & $60-130$\\
$n_{\rm H2}$      [$10^3{\rm cm}^{-3}$]                & 7.4/7.4   & 0.6/1.0   & $0.3-100$/$0.3-20$      & $2.0-7.9$\\ 
$N_{\rm H2,beam}$ [$10^{22}\,{\rm cm}^{-2}$]           & 5.7/4.6   & 1.3/1.0   & $0.08-5.0$/$0.16-12.6$  & $\sim 10$\\
$M_{\rm mol}$     [$10^8\,M_{\odot}$]                  & 1.7/1.4   & 2.0\tablenotemark{f,g} & & $1-10$\tablenotemark{g}\\
\enddata
\tablenotetext{a}{high-excitation component of a two-component LVG model assuming a plane-parallel geometry, using all available CO and $^{13}$CO lines; 
for a spherical model, $n_{\rm H2}$ $\sim5$ times larger, $N_{\rm H2}$ $\sim2$ times larger and $T_{\rm kin}$ better constrained and slightly lower \citep{ward03}}
\tablenotetext{b}{median likelihood of possible values, not a self-consistent solution}
\tablenotetext{c}{full range of possible values (95\% confidence interval)}
\tablenotetext{d}{one-component LVG model assuming a spherically symmetric cloud geometry, using only the high-excitation lines CO($7 \to 6$), CO($4 \to 3$), 
CO($3 \to 2$), and $^{13}$CO($3 \to 2$) \citep{mao00}}
\tablenotetext{e}{FWHM, derived from CO($2 \to 1$) and CO($1 \to 0$) spectra}
\tablenotetext{f}{using the average CO column density derived from the 
integrated CO($6 \to 5$) intensity map}
\tablenotetext{g}{scaled to a distance of $D=3.9$\,Mpc, assumed to be the total mass}
\end{deluxetable}

\clearpage

\begin{deluxetable}{lcc}
\tabletypesize{\small}
\tablewidth{0pt}
\tablecaption{Derived integrated intensities of the emission lines at 
218.5\,GHz at 11\arcsec\ resolution. \label{lvgfits}}   
\tablehead{
    & \colhead{NE lobe} & \colhead{SW lobe}
}
\startdata                        
$\int{T_{\rm mb}\,dv}$(\htco($3_{22} \to 2_{21}))_{\rm models}$ [K\,\kms]  & 1.07 & 0.84 \\
$\int{T_{\rm mb}\,dv}$(CH$_3$OH($4_2 \to 3_1\, E))_{\rm fitted}$ [K\,\kms] & 0.61 & 0.10 \\
$3\sigma$ [K\,\kms]                                                        & 0.26 & 0.27 \\
residual noise level at 218.5\,GHz [mK]                                    & 3.0\tablenotemark{a} & 2.3 \\
rms noise level over the whole baseline [mK]                               & 2.4 & 2.7 \\
\enddata
\tablenotetext{a}{for comparison, the residual noise level at the position of 
the  \htco($3_{03} \to 2_{02}$) line is 3.3\,mK}
\end{deluxetable}

\clearpage

\clearpage

\clearpage

\clearpage

\clearpage

\clearpage

\end{document}